\documentclass[11pt]{article}
\usepackage{putex}
\usepackage{amsfonts}
\usepackage{amssymb}
\usepackage{amsmath}
\usepackage[bookmarks = false, colorlinks = true, linkcolor = blue, citecolor = purple]{hyperref}
\usepackage{ulem}
\usepackage{tipa}
\usepackage{graphics,graphicx}
\usepackage{setspace}
\numberwithin{equation}{section}

\newcommand{\HH}{\mathcal{H}}

\newcommand{\EE}{\mathcal{E}}

\newcommand{\ZZ}{\mathcal{Z}}

\newcommand{\AAA}{\mathcal{A}}

\newcommand{\GG}{\mathcal{G}}
\newcommand{\Tr}{\mbox{Tr}}

\newcommand{\LL}{\mathcal{L}}
\newcommand{\OO}{\mathcal{O}}

\newcommand\be{\begin{equation}}
\newcommand\ba{\begin{eqnarray}}
\newcommand\ee{\end{equation}}
\newcommand\ea{\end{eqnarray}}

\definecolor{purple}{rgb}{0.7,0.0,0.5}
\definecolor{huh}{rgb}{0.0,0.6,0.8}
\definecolor{orange}{rgb}{1,0.5,0}
\definecolor{pink}{rgb}{1,0.4,0.4}
\definecolor{light-gray}{gray}{0.75}

\begin{document}

\title{Ryu-Takayanagi Area as an Entanglement Edge Term}
\authors{Jennifer Lin}
\institution{IAS}{{School of Natural Sciences, Institute for Advanced Study, Princeton, NJ, USA}}

\abstract{
By comparing entanglement in emergent gauge theories to the bulk in AdS/CFT, I suggest that the Ryu-Takayanagi area term is an entanglement edge term related to a natural measure on the gauge group. The main technical result in this paper is an argument why the ``extended Hilbert space" definition of entanglement entropy in a lattice gauge theory is applicable to an emergent gauge theory.}
\maketitle

\newpage

\section{Introduction} 
In recent years, it has been suggested that ``spacetime emerges from quantum entanglement" of some pre-geometric degrees of freedom \cite{VanRaamsdonk:2010pw, Maldacena:2013xja}. The most precise realization of this statement to date is the Ryu-Takayanagi formula in AdS/CFT \cite{Ryu:2006bv}, that relates the entanglement entropy (EE) of a spatial subregion in a holographic CFT to the area of the minimal homologous surface in the bulk: 
\be\label{rt}
S_{EE} = \frac{A_{min}}{4G_N} + \mathcal{O}(G_N^0)\,.
\ee
\eqref{rt} reduces to the Bekenstein-Hawking formula in some \cite{Maldacena:2001kr} (though not all \cite{Headrick:2014cta}) cases. With it, one can derive some nice results in AdS/CFT such as the equality of the linearized Einstein equations and the entanglement first law around vacuum AdS \cite{Lashkari:2013koa}, and entanglement wedge reconstruction \cite{Dong:2016eik, Harlow:2016vwg}, which quantifies how nonlocal the support of a local bulk operator is on the boundary.

The progress so far raises the natural question: what is the microscopic meaning of the Ryu-Takayanagi formula (and its cousin, the Bekenstein-Hawking entropy)? Can we understand what they are counting from the bulk point of view? %(For example, can we find an $\alpha'$-exact formula for the RT entropy on the worldsheet, that reduces to $``A/4G_N"$ in the Einstein gravity limit?)

In this note, I will make an analogy between emergent gauge theories and the bulk in AdS/CFT.
The main point is that when one compares EE in an emergent nonabelian gauge theory to the Ryu-Takayanagi formula with the $O(G_N^0)$ FLM correction \cite{Faulkner:2013ana}, the area term $``A/4G_N"$ looks like the gravity analog of an edge term in the EE of the gauge theory that I will define below. 
By ``the EE of an emergent gauge theory" I mean the following. Suppose that we have an explicitly UV-regulated (e.g. lattice-regulated \footnote{As should become clear below, the specific choice of the lattice regulator is not so important, but I will use it to compare to the literature on EE in gauge theories. One does need to choose {\it some} UV regulator since we are interested in non-universal terms in the EE. In quantum gravity, the theory itself should provide a physical UV regulator.}) 
 QFT with a factorizable Hilbert space which is isomorphic to the Hilbert space of a lattice gauge theory in a low-energy subspace. Then one can write the {non-universal, UV-exact} EE of a region as the algebraic EE of gauge-invariant operators in the region plus an edge term. \footnote{Up to a state-independent constant, as will be explained below.} All of this will be explained more precisely below.
  
 Related remarks have appeared elsewhere.
Ref. \cite{Harlow:2016vwg} explained that the RT area operator is in the center of the algebra associated with the entanglement wedge in bulk effective field theory, suggesting that it is an edge mode. The point of this note is to clarify which one it is and to highlight the UV interpretation in the gauge theory analog.  Ref. \cite{Donnelly:2016auv} first pointed out that the Bekenstein-Hawking area resembles the $``\log \dim R"$-type edge term that one finds when computing EE in a lattice gauge theory, using the definition of \cite{Donnelly:2011hn, Donnelly:2014gva}, which I will review below. The main points added here are that the definition in  \cite{Donnelly:2011hn, Donnelly:2014gva} is (up to fine-print) a physical definition of the EE in a UV-regulated emergent gauge theory, with the UV Hilbert space replacing the extended Hilbert space, and that the bulk effective field theory limit of AdS/CFT resembles an emergent gauge theory with the CFT as the UV theory. Hence we replace the B-H with the RT area term, to sharpen their conjecture.

This note is organized as follows. 
In section \ref{s21}, I review the extended Hilbert space definition of EE in lattice gauge theories \cite{Donnelly:2011hn, Donnelly:2014gva}. In section \ref{s22}, I argue that this definition applies to a UV-regulated emergent gauge theory (up to a constant). 
In section \ref{s3}, I compare the emergent gauge theory to AdS/CFT. 
Finally in section \ref{s4}, I speculate about how to interpret the edge term. 
Appendix \ref{a1} compares the extended Hilbert space definition to the algebraic definition of EE in lattice gauge theories \cite{Casini:2013rba}. Appendix \ref{a2} reviews some examples of emergent gauge theories.

\vspace{2mm}

{\bf Note added in v3}: Many of the speculations in section \ref{s4} reflect early confusions, and although I've left the section as is, it should not be read seriously! See the more recent work for a better discussion. The other sections are unaffected.

\section{Entanglement entropy in emergent gauge theories}\label{s2}

\subsection{Extended Hilbert space definition of EE in a lattice gauge theory}\label{s21}

In this section, I will review a formal proposal for how to define EE in a gauge theory. Later on, I will explain why it agrees with the EE in an emergent gauge theory.

Traditionally, when we study EE, we assume that the Hilbert space factorizes. Then we trace out part of it and take the EE to be the von Neumann entropy of the reduced density matrix.
In a gauge theory, the Hilbert space doesn't factorize, so one must do something different. This has motivated several proposals for how to define EE of a lattice gauge theory in recent years. The one that I will review here was suggested by Donnelly \cite{Donnelly:2011hn, Donnelly:2014gva}. The presentation in this section closely follows \cite{Donnelly:2014gva}. Other definitions are reviewed later on in the file, particularly that of \cite{Ghosh:2015iwa, Soni:2015yga, Pretko:2015zva} in section \ref{s221}, and the algebraic definition of Casini et al. \cite{Casini:2013rba} in Appendix \ref{a1}. 

%\footnote{Alternatively, one might complain that there is no reason to expect the EE of a gauge theory to be well-defined at all. One response, which is the position taken here, is that one can define EE wrt the UV theory in an emergent gauge theory, and look for an IR definition that reproduces that answer. Another is that in topological gauge theories such as 2d YM and Chern-Simons theory, EE given by the replica trick is unambigiously defined (see refs \cite{Donnelly:2014gva} and \cite{Dong:2008ft} respectively), and we can look for a more general definition that will agree with it.} 

The proposal of \cite{Donnelly:2011hn, Donnelly:2014gva} is the following. In a lattice gauge theory, suppose that we pick out an entangling region $A$ by drawing a boundary $\partial A$ that cuts some lattice links. Let the ``extended Hilbert space" be the minimal Hilbert space that factorizes across $\partial A$,%
\be
\HH \subset \HH_{ext.} = \HH_A \otimes \HH_{\bar A}\,,
\ee
constructed formally by adding a node to the lattice at every intersection with $\partial A$ and not imposing the gauge constraints at the new nodes (see Figure \ref{fig2} below). 
The EE of region $A$ is then defined to be the von Neumann entropy of the reduced density matrix for region $A$ in the extended Hilbert space. When one computes EE's with this definition, one finds various boundary terms, that we can see by example.

\subsubsection{Example 1: 2d Electrodynamics on ${\bf S}^1$}

\begin{figure}
\centering
\includegraphics[height=1.5in]{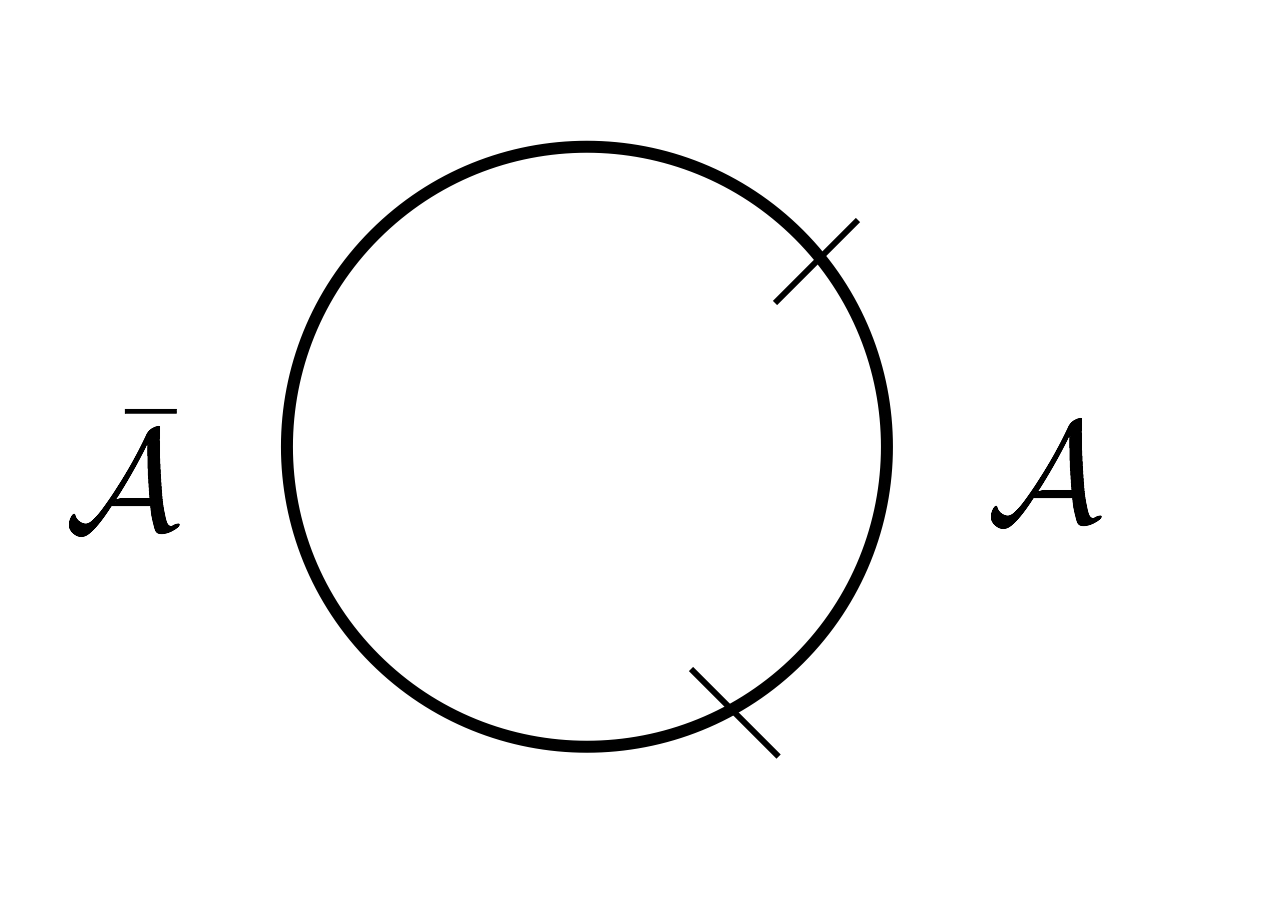}
\caption{EE across an interval on ${\bf S}^1$.}
\label{fig1}
\end{figure}

First, consider a $U(1)$ gauge theory on spatial ${\bf S}^1$. The gauge-invariant operator algebra has only one canonically conjugate pair, the holonomy $\oint {\bf A}$ and the electric field $E(x)$, which is constant everywhere by Gauss's law.  A convenient basis for the Hilbert space are the electric field eigenstates, which are quantized because the gauge group is compact: $\HH : \{|n\rangle\}$ with $E|n\rangle = n|n\rangle$ for $n \in \mathbb{Z}$.

To compute the EE of an interval in the extended Hilbert space prescription, we're instructed to embed $\HH$ into the minimal larger Hilbert space $\HH_{ext.}$ that factorizes across the interval. 
Formally, this should be done by lifting the gauge symmetry at the endpoints of the interval and gauge-fixing again. In this case, the extended Hilbert space doubles the physical one,
\be
\HH_{ext.} = \HH \otimes \HH\,,
\ee
with the unique embedding of states $|n\rangle \rightarrow |n\rangle \otimes |n\rangle$\,.
 An intuitive way to understand the outcome is that for the Hilbert space to factorize, we must be able to cut all extended operators that cross the entangling surface -- in this case, the unique Wilson loop operator -- by adding charges that allow it to break into a pair of Wilson lines. 
 
 We now take the most general state and compute the EE. For 
\be
|\psi\rangle = \sum_n \psi_n |n\rangle \in \HH = \sum_n \psi_n |n\rangle \otimes |n\rangle \in \HH_{ext.}\,,
\ee
 the reduced density matrix in $\HH_{ext.}$ is the diagonal probability distribution over the electric field eigenstates,
\be
\rho_A = \sum_n p_n |n\rangle\langle n|\,, \qquad p_n = |\psi_n|^2\,.
\ee
Hence,
\be\label{shannonee}
S_{EE} = -\sum_n p_n \log p_n\,.\ee

What this ``EE" is computing is the perfect correlation of the electric field operator in regions $A$ and $\bar A$ due to Gauss's law. This type of entropy, that measures kinematic correlations of gauge-invariant operators, is called a ``Shannon edge mode". 

\subsubsection{Example 2: 2d Yang-Mills on ${\bf S}^1$}\label{s212}

Now consider Yang-Mills with gauge group $G$ on the ${\bf S}^1$. The gauge-invariant algebra now contains Wilson loops in all representations, $\Tr_R \exp(i \oint {\bf A})$, and all Casimirs $E^aE^a \dots$ built out of the electric field. One can show (see e.g. \cite{Cordes:1994fc}) that 
%the Hilbert space consists of class functions on the group manifold ($\psi$'s s.t. $\psi(gug^{-1}) = \psi(u)$), and that 
a convenient basis for the Hilbert space, consisting of class functions on the group manifold, is labeled by representations $R$ of $G$: $\HH  : \{|R\rangle\}$\,. 

To compute the EE of an interval, we again extend the Hilbert space by formally lifting the gauge symmetry at the endpoints of the interval and gauge-fixing again. 
Intuitively, we now need to add charges in every representation to cut the loop operators in every representation. This leads to a much larger extended Hilbert space than in the abelian case, with a subspace of size $(\dim R)^2$ (the size of the matrix in each representation) assigned to each state $|R\rangle$:
\be\label{ymext}
\HH_{ext.} = \oplus_R \{|R,i,j\rangle\} \otimes \{|R,i,j\rangle\}, \qquad i,j \in 1, \dots, \dim R\,.
\ee
The unique embedding of the physical state $|R\rangle$ into $\HH_{ext.}$ is 

\be\label{eym}
|R\rangle \quad \rightarrow \quad |R,i,j\rangle \otimes |R,j,i\rangle\,.
\ee

Now for the most general state in the physical Hilbert space, 
\be\label{yms}
|\psi\rangle = \sum_R \psi_R|R\rangle \in \HH = \sum_R\psi_R |R,i,j\rangle \otimes |R,j,i\rangle \in \HH_{ext.}\,,
\ee
the normalized reduced density matrix is
\be\label{ymrho}
\rho_A = \sum_R p_R (\dim R)^{-2} \sum_{i,j} |R,i,j\rangle\langle R,j,i|\,, \qquad p_R = |\psi_R|^2\,,
\ee
and one finds
\be\label{ymee}
S_{EE} = -\sum_R p_R \log p_R + 2\sum_R p_R \log \dim R\,.
\ee

The Shannon edge term appears with the same interpretation as before, but there is a new term specific to the nonabelian case, the ``$\log \dim  R$" edge term. It counts the perfect correlation of surface charges along the boundary in order to make a state in the physical Hilbert space, when the dimensions of representations are greater than 1.

\subsubsection{Higher-dimensional lattice gauge theory}

In a lattice gauge theory in $d >2$ dimensions,
one assigns the Hilbert space for an interval on ${\bf S}^1$ to each lattice link, and 
the gauge-invariant Hilbert space is the tensor product of the Hilbert spaces on the links, modded out by a Gauss constraint at sites. Schematically,
\be\label{hlgt}
\HH_{link} = \oplus_R \{|R,i,j\rangle\}, \quad i,j \in 1, \dots, \dim R\,; \qquad \HH = \frac{\otimes \HH_{link}}{\mbox{Gauss}}\,.
\ee
At each site, the Gauss constraint is implemented by demanding that a Gauss operator $\GG$, acting on the Hilbert spaces of the adjacent links, acts as the identity.
The details are discussed in e.g. \cite{Kogut:1974ag} but are not too important here.

To compute the EE for a subregion $A$ of the lattice whose boundary $\partial A$ intersects a collection of links $\{e\}$, we extend the Hilbert space at each intersection of $\partial A$ with a link. The EE that one gets from this procedure will be the sum of the two edge terms discussed previously, generalized to receive contributions from all boundary links, along with quantum entanglement between interior degrees of freedom that the 2d example is too simple to support. Letting $R_{\partial}$ be the vector of representations labeling the state for all boundary links, %with each $R_{\partial}$ specifying a superselection sector, 
one finds
\be\label{lgtee}
S_{EE} = -\sum_{R_\partial}p_{R_\partial}\log p_{R_\partial} + \sum_{R_\partial}p_{R_\partial}\sum_{e\in\{e\}}\log \dim R_e + \mbox{interior EE}\,.
\ee
%(If the lattice gauge theory has matter at the sites, their entanglement contributes to the interior EE.)

\subsubsection{Comments}

Several comments are in order. First, some technical points for completeness:
\begin{itemize}
\item[(*)] The edge terms in \eqref{lgtee} are non-universal. Their universal part may be relevant for entanglement $c$-theorems \cite{Casini:2006es} and as order parameters for phases of matter: e.g. one needs the Shannon term to recover the topological EE of \cite{Kitaev:2005dm, Levin:2006zz}. See \cite{Soni:2015yga, Pretko:2015zva, Radicevic:2014kqa}.
\item[(*)] Although the Shannon term gets a contribution from each boundary link, it is not quite extensive with the area of the entangling surface, due to the fact that the net flux through closed regions is zero. The analog of this in the $(1+1)d$ example is that it does not depend on the number of intervals.
\item[(*)]
The $``\log \dim R"$-type edge term is local to the entangling surface and extensive with area. 
However, it is state-dependent and can appear in physical quantities that depend on the difference of the EE between states, such as the relative entropy.

\item[(*)]  Only the ``interior EE" in \eqref{lgtee} is distillable  \cite{Soni:2015yga}.
\end{itemize}

More importantly, the following three comments are crucial for the story below.

\begin{enumerate}
\item  An alternative definition for EE in a lattice gauge theory is the algebraic definition popularized by Casini et al.  \cite{Casini:2013rba}: see Appendix \ref{a1} for a review. It turns out that the algebraic entanglement entropy of the maximal gauge-invariant subalgebra supported on a collection of links differs from the extended Hilbert space one \footnote{Where we define an extended Hilbert space for a set of lattice links, instead of an entangling region that cuts through links, to be the lattice Hilbert space with the Gauss law lifted at the boundary sites. See section \ref{s221}.} by the $``\log \dim R"$ edge term \cite{Soni:2015yga}: 
\be\label{algsh}
S_{EE}^{\HH_{ext.}}(\rho_A) = S_{alg, ginv}(A) + \log \dim R \mbox{ edge}\,.
\ee
This is not surprising. While the Shannon EE and interior EE both describe correlations between gauge-invariant operators in regions $A$ and $\bar A$, the $``\log \dim R"$ edge term counts correlations of fictitious, gauge-variant surface charges/Wilson line operators in regions $A$ and $\bar A$, that aren't part of the gauge-invariant operator algebra.
\item From a ``totally IR" point of view, the $``\log \dim R"$ term can be written as the expectation value of a gauge-invariant operator in the center of the operator algebra on regions $A$ and $\bar A$. I.e. there exists some group-dependent $\LL_A$, built from the Casimirs, s.t. $\langle R|\LL_A|R\rangle = \log \dim R$. \footnote{E.g. $\sqrt{4E^aE^a+1}$ for $G=SU(2)$.} 
But this completely obscures the canonical counting interpretation!  The fact that the ``$\log \dim R$" term counts the dimension of a boundary Hilbert space of surface charges is not clear until we introduce the extended Hilbert space, at which point it becomes obvious. See \cite{Kabat:1995jq} for a related discussion.

\item So far, the extended Hilbert space definition of the EE \cite{Donnelly:2011hn, Donnelly:2014gva} is completely formal. 
At this level, the main reason to prefer it to other definitions (e.g. the algebraic one mentioned above) is that both edge terms in \eqref{lgtee} are needed to agree with the replica trick result in topological gauge theories, where one can compute the partition function on replicated manifolds without worrying about the coupling to the conical singularity.  For example, consider the EE of an interval in the de Sitter Hartle-Hawking state of 2d Yang-Mills on ${\bf S}^1$ \cite{Donnelly:2014gva}. This state corresponds to a particular set of coefficients $\psi(R)$ in \eqref{yms}, that one can plug into \eqref{ymrho}, \eqref{ymee}. On the other hand, one can compute the EE by the replica trick, where $\Tr\rho_A$ is $Z_{{\bf S}^2}$, and the EE is a derivative wrt the area of the ${\bf S}^2$, since 2d YM is a TQFT. The answers agree only when both edge terms are included. \\
In fact, for a gauge theory that emerges from a factorizable UV-regulated theory, one can write the UV-exact EE in the form \eqref{lgtee}, up to a state-independent constant.  \eqref{lgtee} is a ``more IR" way of writing the UV-exact answer. This is the subject of the next section.

\end{enumerate} 

\subsection{EE in emergent gauge theories}\label{s22}

I will take an emergent gauge theory to be a theory whose Hilbert space factorizes (``the UV theory"), and whose Hamiltonian is such that a low-energy subspace of the UV Hilbert space is isomorphic to the Hilbert space of a gauge theory. In particular, 
\begin{enumerate}
\item All operators of the gauge theory can be identified with operators of the UV theory s.t. their actions on states respects the duality map.
\item There is ``subregion duality with complementary recovery": given region $\EE_A$ of the gauge theory, there is a region $A$ in the UV theory s.t. all gauge-invariant operators with support only on $\EE_A$ can be identified with UV operators with support only on $A$, and all gauge-invariant operators with support only on $\overline{\EE_A}$ can be identified with UV operators with support only on $\bar A$. 
%There may be multiple UV operators that have the same action on the low-energy subspace.
\item  Also, the inverse is true: operators of the gauge theory not fully contained in either $\EE_A$ or $\overline{\EE_A}$ cannot be reconstructed in just $A$ or $\bar A$.  \footnote{Note that this is not one of the assumptions in \cite{Harlow:2016vwg} which we'll later compare to. The analogous assumption in the related argument of \cite{Harlow:2015lma} is that the reconstruction of the Wilson line anchored to both boundaries in the thermofield double state of AdS/CFT has support on both CFT's by the extrapolate dictionary.}
\end{enumerate}
This is morally the same as the usual definition of an emergent gauge theory in terms of a flow from a UV theory to a gauge theory along the RG, with the assumption that the energy scale of corrections to the gauge theory can be made parametrically large by tuning some external parameter. Examples are given in Appendix \ref{a2}.
Also, to discuss
entanglement edge terms, we need to pick an explicit UV regularization, as mentioned above.
I will assume that the theory is lattice-regulated and that a low-energy subspace of it is isomorphic to the Hilbert space of a lattice gauge theory, possibly on a different lattice. This is to facilitate comparison to \eqref{lgtee}, but it isn't really needed; I'll discuss the regulator-independent interpretation below.

Suppose that in the UV theory we're handed a state in the low-energy subspace, and a region $A$. The goal of this section is to argue that, up to a constant that does not depend on the choice of the state, the microscopic EE of region $A$ wrt the UV Hilbert space equals $\eqref{lgtee}$ for region $\EE_A$.

\subsubsection{Example: Lattice gauge theory without gauge constraints}\label{s221}

\begin{figure}
\centering
\includegraphics[height=2in]{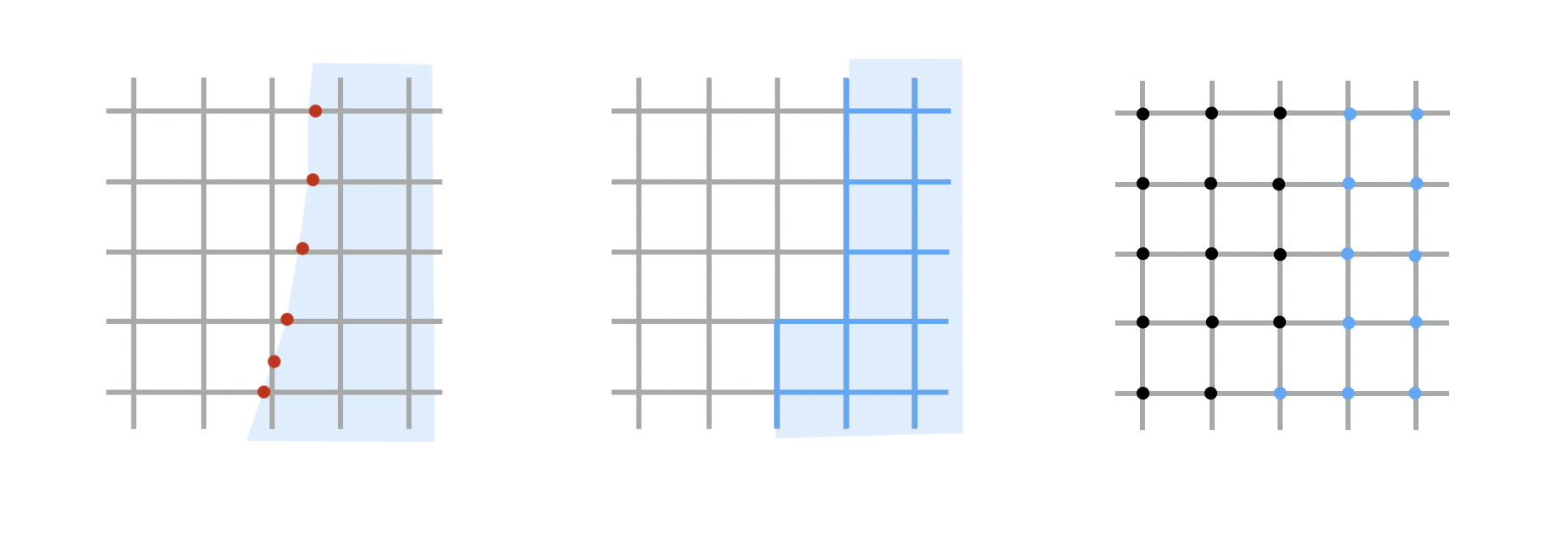}
\caption{The left-hand side illustrates the extended Hilbert space definition of EE in a lattice gauge theory. We take the entangling region $A$ (in blue) to go through a set of links, and extend the Hilbert space at each intersection of $\partial A$ with a link (in red). The middle picture depicts the situation in section \ref{s221}, where we specify an entangling region by a collection of links. The right-hand side illustrates the generic situation in an emergent gauge theory, where the UV Hilbert space might be the tensor product of microscopic Hilbert spaces at the sites of a UV lattice.} \label{fig2}
\end{figure}

As a warm-up, consider the simple class of models where the UV Hilbert space is the Hilbert space of a lattice gauge theory but without imposing gauge constraints at the sites,
\be
\HH = \otimes \HH_{link}
\ee (compare with \eqref{hlgt}), and the Hamiltonian comes with a potential that makes violations of the would-be Gauss law energetically costly. I.e., it contains something like
\be\label{hlink}
H \supset U\sum_i |\GG_i -{\bf 1}|
\ee 
where the index $i$ runs over all lattice sites, $\GG_i$ is the Gauss operator at site $i$, and $U$ is large. In this example, the IR gauge theory lives on the same lattice as the UV theory, and for a region $A = \EE_A$ specified by a collection of lattice links, the UV Hilbert space is almost the extended Hilbert space of region $\EE_A$ in the IR, except that it is much larger away from the entangling surface. This situation was studied in refs \cite{Ghosh:2015iwa, Soni:2015yga, Pretko:2015zva}.

To prove the claim for this example, we must show that the EE does not care that the UV Hilbert space is larger than the extended Hilbert space in the interiors of regions $A$ and $\bar A$. 
But this is true because we can choose a basis for the interiors of regions $A$ and $\bar A$ that split $\HH_A$ and $\HH_{\bar A}$ into a direct sum of subspaces satisfying the emergent gauge constraint at interior lattice sites and not, \footnote{
To do this in practice, we would reconstruct all the Gauss operators of the lattice gauge theory in the interiors of regions $\EE_A, \overline{\EE_A}$ and use their UV images to pick bases for $A$ and $\bar A$. 
}
 and the latter sectors of the Hilbert space cannot contribute to the trace in either $\Tr_{\bar A}\rho$ or $\Tr_A \rho_A \log \rho_A$, by the assumption that the initial state was in the low-energy subspace. 
 
 %  Note that we can do this in any emergent gauge theory provided that we can identify and reconstruct the Gauss operators that annihilate physical states in the IR, and use their UV images to pick bases for $A$ and $\bar A$ as a direct sum of subspaces that do and do not satisfy the constraints. If we imagine a more complicated theory where an emergent IR constraint is $k$-local, the edge terms would accordingly be smeared out over a $k$-site boundary layer. In the analogy to AdS/CFT below, this may be relevant for the RT formula at higher orders in $1/N$, e.g. in \cite{Engelhardt:2014gca}.

\subsubsection{General case}

In general, the UV and IR theories will be less obviously related, e.g. not living on the same lattice. Nonetheless, we now argue that the UV-exact EE takes the form \eqref{lgtee} up to a state-independent constant.

First, note that the extended Hilbert space of section \ref{s21} is a representation of a formal extended operator algebra that contains, in addition to the gauge-invariant operator algebra, Wilson lines in every representation that end on the entangling surface. As mentioned above, the ``$\log \dim R$" edge term comes from the correlations of these Wilson lines in a gauge-invariant state. \footnote{It may be helpful to consider how this works in the previous example \ref{s221}. Imagine building up a gauge-invariant state in the lattice gauge theory by acting with Wilson loops on a ground state with zero entanglement. Each loop operator is defined by tracing over the holonomies $U_{ij}$ on the associated links, which are physical operators in the UV. The indices of the $U_{ij}$'s will be maximally entangled across boundary sites by matrix multiplication, leading to the $``\dim R"$ degeneracy.} From this point of view, the extended Hilbert space definition of the EE for a region $A$ is the algebraic EE of the maximal subalgebra supported on $A$, including the fictitious Wilson lines.

%{\it Relating the extended Hilbert space to an extended operator algebra.} We first observe that, as mentioned above, the extended Hilbert space of section \ref{s21} can be thought of as a representation of a formal extended operator algebra, that contains in addition to the gauge-invariant operator algebra, Wilson lines in every representation ending on the entangling surface. In particular, the $``\log \dim R"$ edge term comes from the correlation of the Wilson lines in a gauge-invariant state.

On the other hand, in an emergent gauge theory, such Wilson lines are not so fictitious. This argument was made in \cite{Harlow:2015lma}. Since the UV Hilbert space factorizes, we must be able to write the UV image of any operator with support on both $\EE_A$ and $\overline{\EE}_A$, including the UV reconstruction of a Wilson loop in any representation, as a sum of tensor products of operators in $A$ and $\bar A$.
This means that in the UV (or at some intermediate scale), there are charged fields to cut the Wilson loop, and 
the extended Hilbert space of the lattice gauge theory is a subspace of the UV Hilbert space, with an isomorphism between all states and operators. (Of course, the UV Hilbert space is generally much larger than the extended Hilbert space.)

% {\it Wilson loops factorize in emergent gauge theories.} The second point is that in an emergent gauge theory, such Wilson lines are not so fictitious; this is the argument of \cite{Harlow:2015lma}. Since the UV Hilbert space factorizes, we must be able to write any operator, including the UV reconstruction of a Wilson loop in any representation with support on both $\EE_A$ and $\overline{\EE}_A$, as a sum of tensor products of operators in $A$ and $\bar A$. This means that in the UV (or at some intermediate scale) there are charged fields to cut the Wilson loop.
%I.e. factorizability in the UV implies that the extended Hilbert space of the lattice gauge theory, 
% the minimal Hilbert space that factorizes
% not just the physical Hilbert space, is equivalent to a subspace of the UV Hilbert space, with an isomorphism between all states and operators. Of course, the UV Hilbert space is generally much larger than the extended Hilbert space.

We want to show that the EE of the maximal subalgebra on region $A$ in the UV equals the EE of the maximal ``extended operator subalgebra" on region $\mathcal{E}_A$, up to a state-independent constant. By the assumption that we start with a state in the low-energy subspace, the reduced density matrix whose von Neumann entropy gives the algebraic EE of region $A$ in the UV is the image of the reduced density matrix of region $\EE_A$ in the IR extended Hilbert space, with zero support on other UV operators (generalizing the argument in \ref{s221}). But now we are comparing the EE's of isomorphic subalgebras, which can differ only by a constant from the sizes of the representations of the algebras.

% {\it EE's of isomorphic subalgebras are equal up to a constant.} We wanted to show that the EE of region $A$ in the UV, algebraically the EE of the maximal subalgebra on region $A$, is the same as the EE of region $\EE_A$ in the extended Hilbert space of the IR gauge theory, algebraically the maximal ``extended operator subalgebra" on region $\EE_A$, up to a state-independent constant.

%\orange{Do we need/want complementary recovery? I think this is OK even for situation in the following comments, but check it...} %... The microscopic EE of region $A$ is equal to what one finds from computing EE for region $\EE_A$ in the IR lattice gauge theory, %, for an entangling cut going through links that separate UV sites in $A$ and $\bar A$, 
%using the extended Hilbert space definition \eqref{lgtee},
%with the entangling boundary either going through lattice sites separating $\EE_A$ from $\EE_{\bar A}$ and the Hilbert space extended at those sites, or through buffer links that lie neither in $\EE_A$ or $\EE_{\bar A}$, with the Hilbert space extended at the intersections of the boundary with the links. 

To summarize, in a lattice-regulated emergent gauge theory, \eqref{lgtee} holds up to a state-independent constant $c$. Schematically,
\be\label{s2sum}
S_{EE, UV}(A) =  S_{EE, IR}(\EE_A) + c = \mbox{Shannon edge}\,\, +\,\, \mbox{$\log \dim R$ edge} \,\, + \,\, \mbox{interior EE} + c\,.
\ee

Comments:

\begin{itemize}
\item[(*)] Regarding the state-independent constant in \eqref{s2sum}, one way to see that we need it is that to any UV theory, we can add a lattice-regulated free massive scalar field with mass $m \gg$ the crossover scale from the UV to IR theory. It will not affect the map between the low-energy manifold of the UV Hilbert space and the Hilbert space of the lattice gauge theory. Each state in the low-energy manifold of the UV Hilbert space now simply comes tensored with the ground state of the scalar field. For a given subregion, the EE of this decoupled sector adds a nonzero constant to the EE relative to what it was before. \\ However, such a state-independent constant should be contrasted with the edge terms in \eqref{lgtee}, which are non-universal but state-dependent. E.g. both edge terms will in general affect the relative entropy of states on the low-energy manifold, while state-independent constants will not.
% % A related, but different issue to the one described above, is how the edge terms in \eqref{lgtee} change under real-space RG as we change the lattice spacing. I will come back to this below.

\item[(*)] The edge terms in \eqref{s2sum} were defined for a lattice regularization, but the argument did not really depend on this. To summarize, in a UV-finite theory with emergent extended objects, the UV-exact EE can be written in a more IR way, as an entanglement entropy assigned to the extended objects fully contained in each region, plus a boundary term that counts the UV degrees of freedom made visible when the IR-extended objects are cut by the entangling surface. 
These UV degrees of freedom are not accessible to operations in the low-energy Hilbert space. All this is precise in compact gauge theory.

% More precisely: the log d_R term is the log of the number of boundary states. This makes it a boundary entropy.

\end{itemize}

\section{Analogy to AdS/CFT}\label{s3}

AdS/CFT is an emergent gauge theory with the CFT as the factorizable UV theory, \footnote{As generally assumed in applications of the RT formula. I stress that we are making an analogy between the IR-emergent gauge theory and the bulk, and we are not interested in the edge terms arising from the gauge symmetry of the CFT.} and bulk effective field theory on AdS, including perturbative gravity, as the IR theory in a low energy subspace. Let us compare the Ryu-Takayanagi formula with its first subleading correction \cite{Faulkner:2013ana} to \eqref{s2sum}. It is not at all obvious that the argument of the previous section, stated in terms of factorizability of Wilson loop operators, carries over to emergent gravity, so this section describes a suggestive analogy and not a proof.
 
% Another assumption, mentioned earlier, is that to write the Hilbert spaces of regions $A$ $(\bar A)$ as a direct sum of states satisfying the gauge constraint and not, we would in practice use the UV images of IR Hamiltonian constraints to carry out the decomposition. We don't know how to do this in AdS/CFT, though perhaps we can roughly divide the Hilbert spaces of regions $A$ $(\bar A)$ into those acted only by only light operators in the CFT and otherwise.
  
For this purpose, a recent repackaging of the RT formula + $1/N$ correction by Harlow \cite{Harlow:2016vwg} is convenient as it separates out the algebraic EE of gauge-invariant bulk operators in the entanglement wedge from the rest of the RT formula. The punchline of \cite{Harlow:2016vwg} is that the RT formula + $1/N$ correction is equivalent to entanglement wedge reconstruction;  \footnote{A third statement, an algebraic version of ``bulk relative entropy = boundary relative entropy" \cite{Jafferis:2015del}, was also shown to be equivalent, but we won't need to use it here.} this was previously shown in \cite{Dong:2016eik}, so 
our comparison does not rely on \cite{Harlow:2016vwg}, but it makes our comparison more straightforward.

Ref. \cite{Harlow:2016vwg} proves the following theorem for all quantum systems.
Suppose that we have a (finite-dimensional) Hilbert space that factorizes, $\HH = \HH_A \otimes \HH_{\bar A}$; a subspace $\HH_{IR} \subseteq \HH$, and a subalgebra $\AAA_{ginv}$ of the operator algebra, whose action on states in $\HH_{IR}$ keeps us inside $\HH_{IR}$. Then, the following statements are equivalent:

\begin{enumerate}
\item There is a subalgebra $\AAA_{ginv,A} \subset \AAA_{ginv}$ s.t. $\forall$ $|\tilde\psi\rangle \in \HH_{IR}$ and $\forall$ $\tilde\OO \in \AAA_{ginv, A}$, there exists an operator $\OO_A$ acting on $\HH_A$ s.t. $\OO_A|\tilde\psi\rangle = \tilde \OO|\tilde\psi\rangle$. Likewise, for all operators in the commutant of $\AAA_{ginv, A}$ on $\HH_{IR}$, $\AAA_{ginv, \bar A}$, there exists an operator supported on $\HH_{\bar A}$ that reproduces its action on $\HH_{IR}$.
\item There exists an operator $\LL_A$ in $\AAA_{ginv, A}\cap \AAA_{ginv, \bar A}$ s.t. $\forall$ $\rho \in \HH_{IR}$, 
\be\label{harlowrt}
S_{EE}(\rho_A) = \Tr(\rho\LL_A) + S_{alg}(\rho, \AAA_{ginv, A})\,,
\ee
where $S_{alg}(\rho, \AAA_{ginv, A})$ is the algebraic EE of $\AAA_{ginv, A}$ in the state $\rho$, as defined in Appendix \ref{a1}.
\end{enumerate}

To interpret this in AdS/CFT, we take $\HH$ to be the CFT Hilbert space, $\HH_{IR}$ to be the low-energy subspace of effective field theory on AdS (or whatever one chooses as the ``code subspace"), $\AAA_{ginv}$ to be the gauge-invariant operators of bulk effective field theory,   and $\AAA_{ginv, A}$ $(\AAA_{ginv, \bar{A}})$ to be the operators of bulk effective field theory with support entirely on the bulk entanglement wedge $\EE_{A}$ $(\EE_{\bar A})$ of boundary regions $A$ ($\bar A$).
Then statement 1 is entanglement wedge reconstruction with complementary recovery \cite{Dong:2016eik}, which identifies region $A$ $(\bar A)$  of the CFT with the entanglement wedge of the bulk EFT on AdS (and its complement),
and statement 2 looks like the Ryu-Takayanagi formula with the $1/N$ correction \cite{Faulkner:2013ana}, where 
\be
\LL_A = \frac{A}{4G_N} + \dots\,,
\ee
the ellipses contain some of the ``Wald-like terms" of the $1/N$ correction \cite{Faulkner:2013ana}, and $S_{alg}(\rho, \AAA_{ginv, A})$ contains the other ``Wald-like terms" as well as $``S_{bulk-ent}"$ of the $1/N$ correction \cite{Faulkner:2013ana}.

If we assume that the ``more IR" formula \eqref{s2sum} for EE in an emergent gauge theory can be used here, then comparing eqs.  \eqref{algsh}, \eqref{s2sum},
to \eqref{harlowrt}, we conclude that the RT area term is a $``\log \dim R"$-type edge term for the bulk. This observation is the main point of this note.

Comments: 

\begin{itemize}
\item[(*)] Eq. \eqref{s2sum} was ambiguous up to a state-independent constant. However, the RT area is state-dependent with a sufficiently large code subspace, so this does not pollute the identification of the RT area term with the $``\log \dim R$"-type edge term. \footnote{Alternatively, one can evade the appearance of state-independent constants in section \ref{s2} by assuming that the crossover scale for the emergence of the gauge theory is higher than all other scales in the problem. In AdS/CFT, gravity presumably emerges at the cutoff scale for the effective field theory, although it may emerge simultaneously with all gauge fields.}
\item[(*)] In the gauge theory example, the $``\log \dim R$" term literally counted the dimensions of the gauge group representations that labeled the links of the lattice intersecting the entangling surface.
Accordingly, the identification suggests (as pointed out by \cite{Donnelly:2016auv}) that perhaps the universal origin of the area term can be understood from the representation theory of the diffeomorphism group. 
\item[(*)] There are related results in the literature. One can study edge terms in tensor networks that resemble AdS/CFT (e.g. \cite{Donnelly:2016qqt}). Previously, at the level of treating 2+1d gravity as a Chern-Simons theory,
% or some quantum version thereof: 2+1d QG ~= quantized SL(2,R) x SL(2,R), the latter factors being copies of Liouville. 
 \cite{Mcgough:2013gka} interpreted the Bekenstein-Hawking entropy of a BTZ black hole as a boundary entropy. 
\item[(*)] As an aside, the bulk Shannon entropy is the entropy of mixing when one considers a state dual to a superposition of classical geometries \cite{Harlow:2016vwg, Almheiri:2016blp}. This fact was not needed for the discussion here, since the Shannon term was absorbed into $S_{alg, ginv}(\mathcal{E}_A)$. 
\item[(*)] The renormalization of $G_N$ \cite{Susskind:1994sm}, and dependence on the choice of the code subspace  of how the bulk EE is distributed between the RT area and the $1/N$ correction \cite{Harlow:2016vwg}, seems to echo the simpler fact that in \eqref{s2sum}, the distribution of the entanglement between the interior EE and edge terms depends on the lattice spacing. It would be interesting to make this analogy precise.
\end{itemize}

\section{Discussion}\label{s4}

To summarize, so far we've argued that the Ryu-Takayanagi area term $``A/4G_N"$ looks like an entanglement edge mode in the bulk, counting correlations of bulk UV degrees of freedom at the entangling surface due to the emergent gauge constraint for gravity. If the RT area is indeed counting UV correlations that are invisible in bulk effective field theory, what is it counting? I conclude with some speculative comments.
\\
\\
{\it Does ``$A/4G_N$" count Chan-Paton factors in string theory?}
% AdS/CFT is actually an emergent perturbative string theory, in the sense that perturbative string theory on a fixed AdS background is a theory of extended objects that omits e.g. the BH microstates on AdS.
To further extend this analogy between emergent gauge theory and the bulk side of AdS/CFT, it is tempting to compare a Wilson loop on the lattice to a closed string in the bulk.
Could it be that at Planckian energies, perturbative closed strings can ``factorize" on the pre-geometric degrees of freedom that make up spacetime, and $``A/4G_N"$ counts the Chan-Paton factors? 

A bit more precisely, $``A/4G_N"$ was identified with the log of the number of boundary states, 
% aka it's a boundary entropy
so would be the number of ways to glue two open strings into a closed one, since the bulk is a string gas. Could it be that the scaling with area comes from the fact that we can do the gluing all along the entangling surface, and the $``O(1/G_N)"$ from there being $O(N)$ Chan-Paton factors at each of the two ends, 
\be
\frac 1{4G_N} \sim \OO(N^2) \sim (\mbox{CP factors})^2?
\ee

\begin{figure}
\centering
\includegraphics[height=2in]{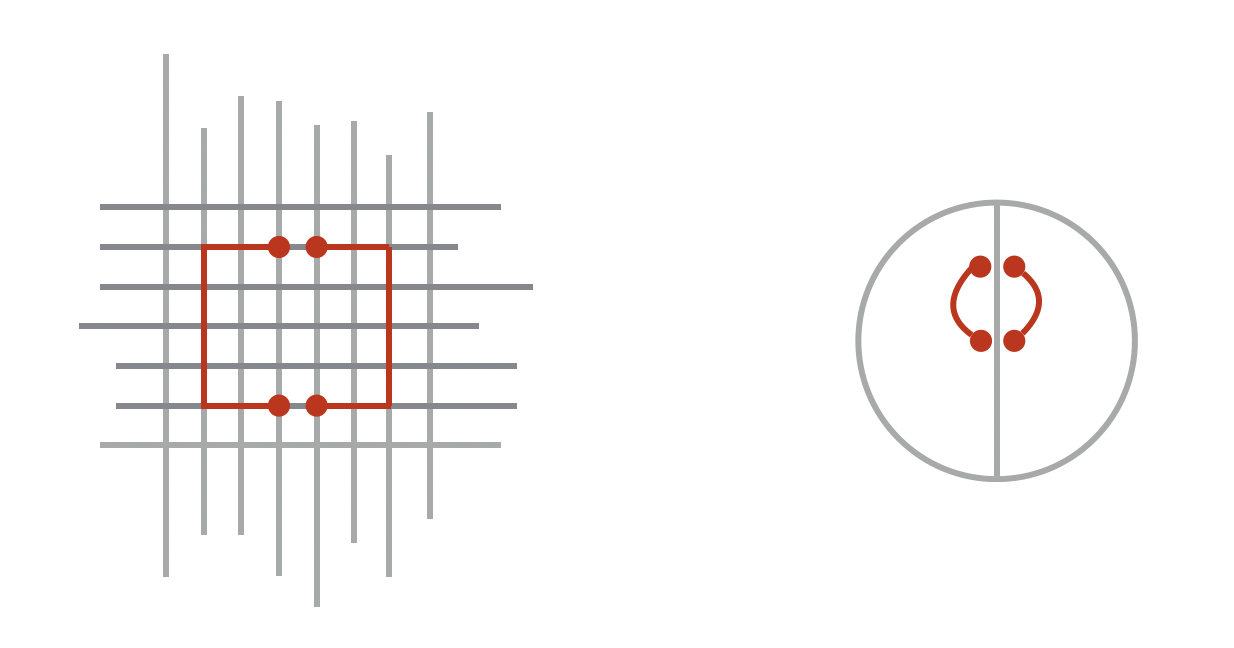}
\caption{ Wilson loop in an emergent gauge theory cut by surface charges on the lattice; closed strings cut by the entangling surface in AdS?}
\label{fig3}
\end{figure}

Perturbative closed string theory, like the theory of hadrons, has a Hagedorn temperature. In the gauge theory, the resolution is that the limiting temperature is a harbinger of a new UV phase, 
% of deconfined quarks/gluons...
in terms of which there is no limiting temperature. By analogy, it has long been suspected that some new non-perturbative phase of string theory describes black holes. The picture that this section suggests is that the black hole microstates are these states where the Chan-Paton indices are liberated. This is like a more Lorentzian version of the Susskind-Uglum story \cite{Susskind:1994sm}.

At face value, these are just ill-defined words. What takes it beyond philosophy is the idea that the entanglement entropy in even the AdS vacuum is a probe of the UV, predominantly counting these open string states at weak gravitational coupling. In this picture, we would see the open strings if we could construct the string dual of a boundary density matrix for a subregion in the vacuum, as discussed below.
%\footnote{At first sight, string theory in AdS$_3$ past the correspondence point \cite{Horowitz:1996nw} poses a challenge for our analogy. On the one hand, it seems to fit into our paradigm where there is a UV theory with a dual description in terms of extended objects; on the other hand, past a critical $g_s \sim N$, string theory on AdS$_3$ does not support black hole states \cite{Giveon:2005mi}. But in this case there is no notion of subregion duality, so our entire discussion doesn't apply.} 
% Tl;dr of above footnote: String theory past the correspondence point is not at weak gravitational coupling.
% BH entropy = log states = 1/G_N = O(N^2) 
% -> BH degeneracy of states = e^entropy = e^{N^2}, it follows from the definition of the BH entropy.

% WRT AdS3 phase diagram: "The hypothesis is that the perturbative closed strings *cannot* fractionate past the correspondence point, which is a phase transition that takes place at a critical gravitational coupling g_S = g_N. We claim that "boundary EE = bulk EE" is dominated by the BH microstates at weak gravitational coupling. As we interpolate past the correspondence point, the BH microstates disappear, so there should be a phase transition in the bulk EE. We know there is no transition in the boundary EE which is always c/3 log L, so presumably the transition is in the replacement of the RT formula by a different bulk prescription.

There is an independent argument that a fundamental string cut by an entangling surface comes with an $O(N)$ degeneracy. We can add a single string to the bulk of empty AdS with a space-filling flavor brane by putting a color singlet quark-antiquark pair at antipodal points of the boundary ${\bf S}^1$ \cite{Jensen:2013ora}. Lewkowycz and Maldacena computed the extra boundary entanglement from the $q\bar q$ pair and found that it always comes with a $``\log N"$, basically from cutting the color flux tube on the boundary \cite{Lewkowycz:2013laa}. But in the bulk, all we did was cut one string with an entangling surface.
\\
\\
{\it What is the string dual of a density matrix?} The entanglement entropy of a region $A$ is just one function of the reduced density matrix, $\rho_A$. If the RT formula can be interpreted as ``boundary EE = bulk EE", a more refined question naturally follows: what is the string dual of the density matrix for a region $A$ in a holographic CFT?  \footnote{Alternatively, the density matrix contains the same information as the entanglement entropy and all the Renyi's. It would be interesting if the ``cosmic brane" in the holographic Renyi entropy formula \cite{Dong:2016fnf} can be interpreted from our point of view. This seems challenging even if a string dual of $\rho_A$ can be defined because the holographic Renyi entropy formula contains a derivative wrt the replica index $n$.} 

The analogy in this section suggests that the Lorentzian string dual of a density matrix $\rho_A$ is the entanglement wedge with some mysterious sector of open strings at the entangling surface in a mixed state of Chan-Paton factors, just as the reduced density matrix in the emergent gauge theory example took the form \eqref{ymrho}. This is consistent with the conclusion of \cite{Czech:2012be} at the level of bulk effective field theory, who argued that the gravity dual of individual Rindler microstates in holographic CFT's are bulk spacetimes that are indistinguishable from the AdS-Rindler wedge except in a region of size $\varepsilon$ around the horizon where geometry breaks down, $\varepsilon$ being the cutoff for the bulk effective field theory.

\begin{figure} 
\centering
\includegraphics[height=2in]{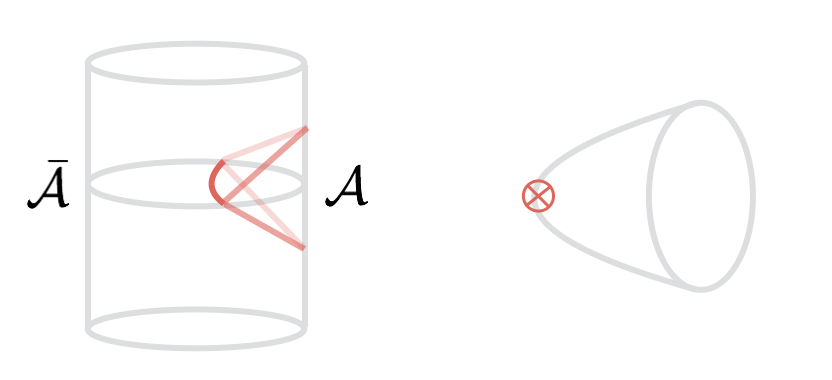}
\caption{Left: The analogy between the Wilson loop and closed string suggests that the string dual of $\rho_A$ is the entanglement wedge with open strings in a mixed state of Chan-Paton factors on a stack of branes at the entangling surface. Right: A naive Euclidean continuation of the left-hand picture is that the perturbative string background whose partition sum reproduces $\log \Tr_A\rho_A$ is the Euclidean cigar with a $T \sim 1/g_s^2$ defect at the tip.}
\label{f5}
\end{figure}

\vspace{4mm}

\noindent {\it A Euclidean version of the conjecture.} If we pick a reduced density matrix in the CFT whose modular Hamiltonian generates a geometric flow, we can ask what happens when we analytically continue the above picture to Euclidean time. For simplicity, let me specialize to a Rindler wedge in the vacuum state of the holographic CFT, whose bulk effective field theory dual at $N=\infty$ is an AdS-Rindler wedge. 

It's tempting to make the following guess. If the Lorentzian dual to a density matrix contains open strings ending on the entangling surface, a naive analytic continuation of the Rindler time on both boundary and bulk suggests that  the Euclidean background whose perturbative string partition sum equals (the unnormalized) $\log\Tr\rho_{A}$ of the boundary CFT is not the smooth cigar, but the cigar with a codimension-2 object of tension $T \sim 1/g_s^2$ at the tip of the cigar. \footnote{
Note that the answer to a superficially related question, ``the QFT partition sum on which Euclidean manifold equals $\Tr\rho_{Rindler}$?" for the Rindler wedge of a QFT on $\mathbb {R}^{d,1}$ (and the generalization to other situations where the modular Hamiltonian generates geometric flow), is the smooth Euclidean manifold without any defect at the origin, by the usual path integral argument. 
Namely, we set up the density operator on region $A$ by a Euclidean path integral with two open cuts along $A$ at $\tau = 0$, and $\Tr\rho_A$ sews together the open cuts.
But it does not follow that the Euclidean string background whose partition function gives $\log\Tr\rho_{Rindler}$ on the boundary must be the smooth cigar. We cannot trace out half a string background as a worldsheet operation, and quantum gravity versions of this argument are only defined holographically, with the smooth cigar being the dominant saddle in Einstein gravity.} 

I want to emphasize that this represents an extra level of speculation from the discussion up to now, and could be wrong even if the Lorentzian picture can be made precise. But there is a recent example consistent with it \cite{Donnelly:2016jet}: 

% On the other hand, it could be that as one makes this analytic continuation, the ends of the open strings "glue" together with themselves, and we are just left with the smooth cigar.

% By definition, $\rho_A$ on the boundary reproduces the expectation values of all CFT operators in region $A$. Thus, string amplitudes in the dual background must agree with the amplitudes in the global space, with absorption and re-emission by the end-of-the-world brane conspiring to produce this result. This is like a string theory version of the membrane paradigm.

2d Yang-Mills on ${\bf S}^1$, at large $N$, looks like a string field theory. One can choose a basis for the Yang-Mills Hilbert space labeled by elements of the symmetric group, and interpret the cycles as closed strings wound around the ${\bf S}^1$ \cite{Baez:1994gk}. The analog of the Rindler wedge for a QFT on ${\bf S}^1$ (whose modular Hamiltonian generates a geometric flow) is an interval $A$ in the $dS_2$ Hartle-Hawking state. For 2d Yang-Mills, the Euclidean partition sum that reproduces $\Tr_A\rho_{A}$ is the Yang-Mills partition sum on the smooth ${\bf S}^2$, as it had to be.   On the other hand, in the ``string field" interpretation, 
the partition sum that reproduces $\Tr_A\rho_{A}$ is not the partition sum on the smooth sphere, but rather, the partition sum on the sphere with two $T\sim 1/g_s^2$ defects at the ends of the interval $A$. 
Namely, one can expand $Z_{{\bf S}^2}^{YM}$ as a sum over branched coverings of the ${\bf S}^2$ that wrap two pointlike defects at the ends of the interval, with a factor of $N$ associated to each closed cycle around the defects \cite{Gross:1993yt}. The two points are the analog of the tip of the cigar in higher-dimensional examples.

An earlier version of this note suggested looking for signatures of a defect at the tip of the cigar in the large $k$ limit of the $SL(2)/U(1)$ CFT \cite{Witten:1991yr}. This was based on the suggestion in \cite{Giveon:2013ica, Giveon:2014hfa} that this CFT comes with extra states relative to the CFT of two free bosons. However, the extra states of \cite{Giveon:2013ica, Giveon:2014hfa} seem to be an artifact of the regularization. See \cite{Giveon:2015raa, Troost:2017fpk} for related comments.
%
%For example, they found that contributions to the elliptic genus of the ($N=2$ version of the) $SL(2)/U(1)$ CFT from normalizable discrete states at finite $k$, survive the large $k$ limit \cite{Giveon:2014hfa}. In \cite{Giveon:2013ica}, they also argued that the contribution from discrete states in the ${\bf T}^2$ partition function of the $N=2$ $SL(2)_k/U(1)$ CFT, as isolated following  \cite{Maldacena:2000kv}, survives the large $k$ limit. However, this computation is more subtle than the elliptic genus one in that the contribution from the continuum sector diverges and can contribute at order $k^0$ as well. Hence it need not follow that the ${\bf T}^2$ partition function of the full theory at large $k$ is different than in flat space. Indeed, with a different regulator, they later found a different  answer \cite{Giveon:2014hfa}. 
%
% NOTE THAT THIS "STRING DUAL OF A DENSITY MATRIX" STUFF IS UN-TESTABLE IN THE ABSENCE OF A GEOMETRIC MODULAR FLOW.
\\
\\
\noindent {\it Can we make any of this precise?} To summarize, in this section we have speculated that quantum gravity factorizes along bulk extremal surfaces; that perturbative closed strings can ``factorize" in the Hilbert space of non-perturbative string theory; and that the string dual of a reduced density matrix for a subregion in a holographic theory, at weak gravitational coupling, contains a mixed state of open strings ending on the boundary of the associated bulk subregion. The last of these statements is the only one that is really well-defined.

In order to show it, we would have to construct the string dual of a boundary density matrix in a holographic system. We certainly can't do this in AdS/CFT, since an obviously easier warm-up problem is to construct the worldsheet CFT for strings in the vacuum from knowing everything about the CFT, which we don't (in general) know how to do.

But there is another string/non-string duality where we know the algorithm how to get the closed string worldsheet CFT about the vacuum from the non-gravity dual. This is the $c=1$ matrix model. Moreover, we have an understanding of bulk locality in the weakly coupled part of the bulk \cite{Hartnoll:2015fca}. Can we use this system to study the conjecture outlined in this section? \footnote{
The reader might complain that the $c=1$ matrix model is qualitatively different from AdS/CFT in that it doesn't contain the very black holes whose microstates we want to count \cite{Karczmarek:2004bw}. Indeed, for the construction envisioned here, we would have to assume the outstanding conjecture (see e.g. \cite{Kazakov:2000pm}) that the $c=1$ matrix model in the double scaling limit is the analog of $N = \infty$ in AdS/CFT, and that black holes are present at large but finite $N$ when the non-singlet states aren't completely gapped out. 
} I hope to report on this in the future.

% But bulk locality is more complicated in the matrix model. In AdS/CFT, bulk locality of light (N=\infty) operators is subregion-subregion. The generalization to heavy (N != \infty) operators is completely obvious. 

% In the matrix model, bulk locality of singlet (N = \infty) operators seems to assign the eigenvalue density in a range of x to a subregion. The generalization to non-singlet operators is less obvious.

\subsection*{Acknowledgments}
I thank Raphael Bousso, Xi Dong, William Donnelly, Dan Harlow, David Kutasov, Juan Maldacena, Mark Mezei, Rob Myers, Djordje Radicevic, Steve Shenker and Gabriel Wong for hepful comments. I am grateful to the Perimeter Institute for hospitality as this work was being completed. My work is supported by the William D. Loughlin Membership at the IAS and by the U.S. Department of Energy.

\begin{appendix}
\section{Algebraic definition of EE in lattice gauge theories}\label{a1}

Given a state $|\psi\rangle$ in the Hilbert space of a quantum system and a subalgebra $\AAA_0$ of the operator algebra, one can define an EE for the subalgebra (see e.g. \cite{Ohya} and a review in \cite{Harlow:2016vwg}). The starting point is that there is in general a unique element $\rho \in \AAA_0$ s.t. \be\label{algee} \Tr_\HH(\rho\OO) = \langle \psi|\OO|\psi\rangle, \qquad \forall \OO \in \AAA_0\,, \ee since one can expand $\rho = \sum_{\OO_i \in \AAA_0} p_i \OO_i$ in a basis for $\AAA_0$, and the condition \eqref{algee} gives one equation for each unknown $p_i$. The von Neumann entropy of $\rho$ is well-defined, and it is tempting to take it to be the EE of the subalgebra, 
\be \label{vne}
S_{EE}(\AAA_0) ?= \Tr_\HH\rho \log \rho\,.
\ee 

However, this concise definition has the shortcoming that if we also take the Hilbert space $\HH$ to be the global one, the EE of the maximal subalgebra on a region $A$ in a factorizable QFT will not equal the EE one obtains from the partial trace, differing by a constant related to the ratio of the dimension of the global Hilbert space to the dimension of the Hilbert subspace on region $A$.  One can always pick out the appropriate representation by hand, or define \eqref{vne} wrt the global Hilbert space and keep this constant in mind for applications, but to automatically land on the standard result when the Hilbert space factorizes, one can proceed by the following algorithm \cite{Casini:2013rba}:

Given a subalgebra $\AAA_0$ with a center $\ZZ$, choose a basis that diagonalizes $\ZZ$ in the global Hilbert space. In this basis, the elements of $\AAA_0$ will take a block-diagonal form, and the algebra generated by $\AAA_0$ and its commutant $\AAA_0'$ will have the form
\be
\left(\begin{tabular}{cccc} $\AAA_1 \otimes \AAA_1'$ & 0 & $\dots$ & 0 \\ 0 & $\AAA_2 \otimes \AAA_2'$ & \dots & 0 \\ \vdots & \vdots & & \vdots \\ 0 & 0 & 0 \dots & $\AAA_m \otimes \AAA_m'$ \end{tabular} \right)\,,
\ee
with each $\AAA_k, \AAA_k'$ included in $\AAA_0, \AAA_0'$ respectively.

When $\ZZ$ is nontrivial, $\AAA_0$ and its commutant do not generate the entire algebra $\AAA$, and the global density matrix $\rho$ may have off-diagonal elements. The algorithm instructs us to erase the off-diagonal elements of $\rho$. Then in each block, we partial trace over $\AAA_k'$. The von Neumann entropy of the resulting density matrix agrees by construction with the standard result in a factorizable theory, with $\AAA_0$ the maximal subalgebra on a factor, and with a trivial center.
%This algorithm still has the feature that one computes the von Neumann entropy of the unique density matrix belonging to the algebra $\AAA_0$ and giving the correct expectation values, but readjusts the size of the Hilbert space in \eqref{vne} to irreducibly represent the subalgebra.
%
This algorithm is what I will refer to as the ``algebraic EE" in this paper.

Casini et al. \cite{Casini:2013rba} suggested to define the EE of a collection of links in lattice gauge theory as the algebraic EE of a gauge-invariant subalgebra supported on the region. Actually, ref. \cite{Casini:2013rba} offered multiple definitions, corresponding to different choices of the subalgebra for a given region.  In their ``electric center" choice, one takes the EE of a collection of links in lattice gauge theory to be the algebraic EE of the maximal gauge-invariant subalgebra supported on the region. 
In their ``magnetic center" choice, one takes the EE on a collection of links to be the algebraic EE of the maximal gauge-invariant subalgebra supported in the {interior} of the region, excluding the boundary links. The claim to fame of the magnetic center choice is that it is related to the electric center under duality (so maximal algebras dualize to non-maximal ones; see e.g. \cite{Radicevic:2016tlt, Donnelly:2016mlc}). In this file, I refer specifically to the electric center choice, or choice of the maximal subalgebra on a region as the ``algebraic definition".

\begin{figure}
\centering
\includegraphics[height=1.8in]{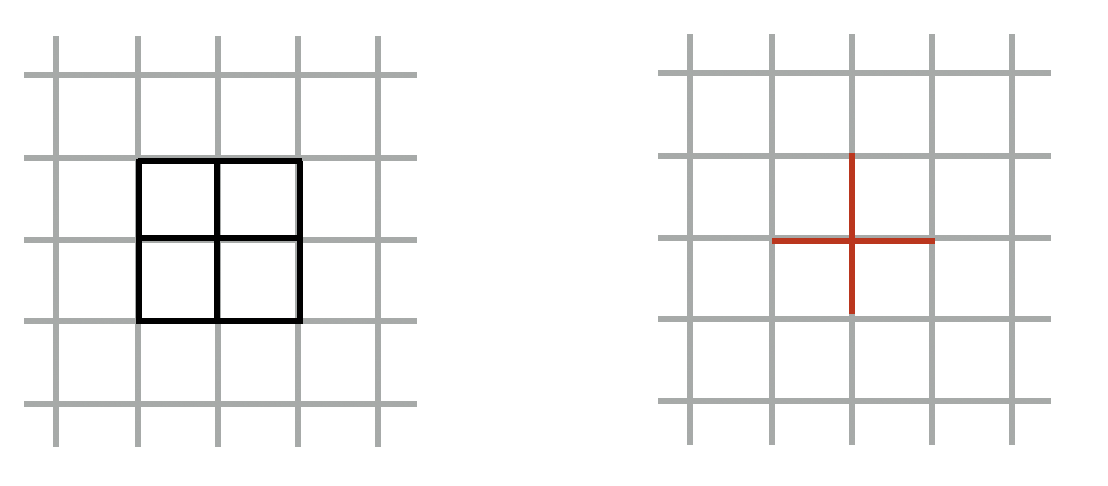}
\caption{The electric center choice of \cite{Casini:2013rba} defines the EE for the collection of links marked in black above as the algebraic EE of the maximal subalgebra supported on those links. The magnetic center choice defines the EE for the collection of links marked in black to be the algebraic EE of the maximal subalgebra supported on the red links.
}
\label{fig6}
\end{figure}

As mentioned above, \eqref{algsh}, the algebraic definition of EE in a lattice gauge theory and the extended Hilbert space definition of section \ref{s21} turn out to disagree in nonabelian gauge theories by the  $``\log \dim R"$ edge term \cite{Soni:2015yga}.  From the point of view of emergent gauge theory, the reason is that the extended Hilbert space definition gives the fine-grained EE WRT a UV observer, while the algebraic EE for gauge-invariant operators is a coarse-grained EE for an observer below the UV/IR crossover scale, that does not see the correlations of the UV degrees of freedom along the entangling edge. From this point of view, the different center choices in \cite{Casini:2013rba} are different coarse-grainings.

\section{Examples of emergent gauge theories}\label{a2}

In this section, I review some examples of emergent gauge theories. This section is basically a brief summary of the references given below.

\subsection{Toric code}\label{a21}

The Kitaev toric code \cite{Kitaev:1997wr} is the simplest example of an emergent gauge theory. It belongs to the class of model discussed in Section \ref{s221}, where the UV Hilbert space is the Hilbert space of a lattice gauge theory without the Gauss constraints at the vertices. Namely, one assigns a qubit to each lattice link, and the UV Hilbert space is the tensor product of the Hilbert spaces on the links, \eqref{hlink}.
The Hamiltonian is %
\be\label{hz2}
H = -U \sum_i \prod \sigma^x - g \sum_p \prod \sigma^z\,,
\ee
where $i$ runs over the lattice sites; $p$ runs over the plaquettes; one takes the product over all links that end on site $i$ in the first term, and over all links in plaquette $p$ in the second term.
Because the two terms in \eqref{hz2} commute, we can diagonalize the Hamiltonian by independently minimizing each.
The Gauss operators in $\mathbb{Z}_2$ lattice gauge theory are $\GG = \prod \sigma^x$ over all links adjacent to a site, with eigenvalue $+1$ on gauge-invariant states and $-1$ otherwise. Hence as we make $U$ parametrically large ($\gg g$ and the lattice spacing), \eqref{hz2} is minimized on gauge-invariant states, and the low-energy Hilbert space of the toric code coincides with that of $\mathbb{Z}_2$ lattice gauge theory. 
%().
% The second term in \eqref{hz2} is the $\mathbb{Z}_2$ lattice analog of $\int d^3x \vec B^2$. 

This model is solvable and has interesting properties outside the scope of the discussion here. 
On topologically nontrivial manifolds, the ground state is degenerate, exhibiting topological order.
The phase diagram of the theory as one tunes the relative strengths of perturbations to the Hamiltonian is understood.
See \cite{Wen:2004ym, McGreevy:2016myw} for further discussion.

% Perturbations can correspond to string tensions and fugacities for the string endpoints; by making these terms in the Hamiltonian large, we reach different phases of the gauge theory, where it is confined or Higgsed. If we insert a pair of charges for large but not infinite g, they are confined by the potential (potential grows linearly with the separation). 

\subsection{Example with continuum limit}\label{a22}

The previous example is somewhat unsatisfactory from the high energy theorist's point of view because there is no continuum version of the UV theory. In continuum examples, the duality map is often less obvious, and one sees the emergent gauge theory in two steps: first a kinematic change of variable makes an auxiliary vector field appear, then one must show that a kinetic term for it is generated by dynamics.
%One can do these steps on the lattice as well.
The latter is easier to see with continuum methods (see e.g. the recent discussion of the $\mathbb{CP}^{N-1}$ model in \cite{Harlow:2015lma}).
Given a priori knowledge that a kinetic term is generated, the change of variable explains how to explicitly ``reconstruct operators" of the IR theory.

Here is an example taken from ref. \cite{Lee:2010fy}.
Consider a 4d Euclidean cubic lattice with $N(N-1)/2$ bosonic quantum variables at each site, each valued on an ${\bf S}^1$ target. Let us label them by $e^{i\theta^{ab}}$ with $\theta^{ab} = -\theta^{ba}$ and $a,b, \in 1, \dots, N$. 
% Presumably there is also some sort of large N limit.
We take the Hamiltonian to be
\be
H = -t\sum_{\langle i,j\rangle}\sum_{a,b} \cos\left(\theta^{ab}_i - \theta^{ab}_j \right) + K\sum_i \sum_{a,b,c} \cos(\theta_i^{ab}+\theta_i^{bc}+\theta_i^{ca})\,,
\ee
where the index $i$ labels lattice sites and $\langle i,j\rangle$ denotes the sum over nearest-neighbor sites.
The first term is the kinetic term for the $\theta^{ab}$'s in the continuum, and the second is a potential for them. 
Now we take the large $K$ limit.
The potential imposes the dynamical constraint \be\label{dcc} \theta_i^{ab} + \theta_i^{bc} + \theta_i^{ca} = 0\,. \ee

Solutions of \eqref{dcc} can be parametrized by new variables
\be\label{meson}
\theta^{ab}_i = \phi^a_i - \phi^b_i\,,
\ee
that are unrestricted on the low-energy manifold.
The effective Hamiltonian on the low energy manifold is 
\be\label{partonh}
H = -t\sum_{\langle i,j\rangle}\left(\sum_a e^{i(\phi^a_i - \phi^a_j)} \right)\left(\sum_b e^{-i(\phi_i^b-\phi_j^b)} \right)\,.
\ee
With another change of variable 
% introducing an auxiliary field for each link ij
%
\be\label{nij}
\eta_{ij} = \sum_b e^{-i(\phi_i^b - \phi_j^b)}\,,
\ee
 \eqref{partonh} becomes
\be
H = t\sum_{\langle i,j\rangle}\left[|\eta_{ij}|^2 - |\eta_{ij}|\sum_a e^{i(\phi^a_i - \phi^a_j - a_{ij})} - cc. \right]
\ee
with $a_{ij}$ the phase of $\eta_{ij}$, which is invariant under
\begin{eqnarray*}
\phi^a_i &\rightarrow & \phi^a_i + \varphi_i\,,\\
a_{ij} &\rightarrow & a_{ij} + \varphi_i - \varphi_j
\end{eqnarray*}
for any $\varphi$.
Hence the low-energy effective Hamiltonian has a local $U(1)$ symmetry. 

Gauge-invariant IR operators, e.g. products of $\eta_{ij}$'s around closed loops on the lattice, can be mapped explicitly to the UV $\theta$'s \eqref{meson}: this is built into the redefinition. Since the gauge theory in this case is $U(1)$, this example is too simple to support a $``\log \dim R$"-type edge term, but illustrates the general idea.

%The bare gauge coupling is infinite, but if one integrates high energy modes of the $\phi$'s, one generates a finite coupling for the Maxwell term of the Hamiltonian in lattice gauge theory; this is the second step.

\end{appendix}

\newpage
\bibliographystyle{ssg}
\bibliography{rtlgt}

\begingroup\raggedright\begin{thebibliography}{10}

\bibitem{VanRaamsdonk:2010pw}
M.~Van~Raamsdonk, ``{Building up spacetime with quantum entanglement},'' {\em
  Gen. Rel. Grav.} {\bf 42} (2010) 2323--2329,
  \href{http://xxx.lanl.gov/abs/1005.3035}{{\tt 1005.3035}}. [Int. J. Mod.
  Phys.D19,2429(2010)].

\bibitem{Maldacena:2013xja}
J.~Maldacena and L.~Susskind, ``{Cool horizons for entangled black holes},''
  {\em Fortsch. Phys.} {\bf 61} (2013) 781--811,
  \href{http://xxx.lanl.gov/abs/1306.0533}{{\tt 1306.0533}}.

\bibitem{Ryu:2006bv}
S.~Ryu and T.~Takayanagi, ``{Holographic derivation of entanglement entropy
  from AdS/CFT},'' {\em Phys. Rev. Lett.} {\bf 96} (2006) 181602,
  \href{http://xxx.lanl.gov/abs/hep-th/0603001}{{\tt hep-th/0603001}}.

\bibitem{Maldacena:2001kr}
J.~M. Maldacena, ``{Eternal black holes in anti-de Sitter},'' {\em JHEP} {\bf
  04} (2003) 021, \href{http://xxx.lanl.gov/abs/hep-th/0106112}{{\tt
  hep-th/0106112}}.

\bibitem{Headrick:2014cta}
M.~Headrick, V.~E. Hubeny, A.~Lawrence, and M.~Rangamani, ``{Causality \&
  holographic entanglement entropy},'' {\em JHEP} {\bf 12} (2014) 162,
  \href{http://xxx.lanl.gov/abs/1408.6300}{{\tt 1408.6300}}.

\bibitem{Lashkari:2013koa}
N.~Lashkari, M.~B. McDermott, and M.~Van~Raamsdonk, ``{Gravitational dynamics
  from entanglement 'thermodynamics'},'' {\em JHEP} {\bf 04} (2014) 195,
  \href{http://xxx.lanl.gov/abs/1308.3716}{{\tt 1308.3716}}.

\bibitem{Dong:2016eik}
X.~Dong, D.~Harlow, and A.~C. Wall, ``{Reconstruction of Bulk Operators within
  the Entanglement Wedge in Gauge-Gravity Duality},'' {\em Phys. Rev. Lett.}
  {\bf 117} (2016), no.~2 021601,
  \href{http://xxx.lanl.gov/abs/1601.05416}{{\tt 1601.05416}}.

\bibitem{Harlow:2016vwg}
D.~Harlow, ``{The Ryu-Takayanagi Formula from Quantum Error Correction},''
  \href{http://xxx.lanl.gov/abs/1607.03901}{{\tt 1607.03901}}.

\bibitem{Faulkner:2013ana}
T.~Faulkner, A.~Lewkowycz, and J.~Maldacena, ``{Quantum corrections to
  holographic entanglement entropy},'' {\em JHEP} {\bf 11} (2013) 074,
  \href{http://xxx.lanl.gov/abs/1307.2892}{{\tt 1307.2892}}.

\bibitem{Donnelly:2016auv}
W.~Donnelly and L.~Freidel, ``{Local subsystems in gauge theory and gravity},''
  {\em JHEP} {\bf 09} (2016) 102,
  \href{http://xxx.lanl.gov/abs/1601.04744}{{\tt 1601.04744}}.

\bibitem{Donnelly:2011hn}
W.~Donnelly, ``{Decomposition of entanglement entropy in lattice gauge
  theory},'' {\em Phys. Rev.} {\bf D85} (2012) 085004,
  \href{http://xxx.lanl.gov/abs/1109.0036}{{\tt 1109.0036}}.

\bibitem{Donnelly:2014gva}
W.~Donnelly, ``{Entanglement entropy and nonabelian gauge symmetry},'' {\em
  Class. Quant. Grav.} {\bf 31} (2014), no.~21 214003,
  \href{http://xxx.lanl.gov/abs/1406.7304}{{\tt 1406.7304}}.

\bibitem{Casini:2013rba}
H.~Casini, M.~Huerta, and J.~A. Rosabal, ``{Remarks on entanglement entropy for
  gauge fields},'' {\em Phys. Rev.} {\bf D89} (2014), no.~8 085012,
  \href{http://xxx.lanl.gov/abs/1312.1183}{{\tt 1312.1183}}.

\bibitem{Ghosh:2015iwa}
S.~Ghosh, R.~M. Soni, and S.~P. Trivedi, ``{On The Entanglement Entropy For
  Gauge Theories},'' {\em JHEP} {\bf 09} (2015) 069,
  \href{http://xxx.lanl.gov/abs/1501.02593}{{\tt 1501.02593}}.

\bibitem{Soni:2015yga}
R.~M. Soni and S.~P. Trivedi, ``{Aspects of Entanglement Entropy for Gauge
  Theories},'' {\em JHEP} {\bf 01} (2016) 136,
  \href{http://xxx.lanl.gov/abs/1510.07455}{{\tt 1510.07455}}.

\bibitem{Pretko:2015zva}
M.~Pretko and T.~Senthil, ``{Entanglement entropy of $U(1)$ quantum spin
  liquids},'' {\em Phys. Rev.} {\bf B94} (2016), no.~12 125112,
  \href{http://xxx.lanl.gov/abs/1510.03863}{{\tt 1510.03863}}.

\bibitem{Cordes:1994fc}
S.~Cordes, G.~W. Moore, and S.~Ramgoolam, ``{Lectures on 2-d Yang-Mills theory,
  equivariant cohomology and topological field theories},'' {\em Nucl. Phys.
  Proc. Suppl.} {\bf 41} (1995) 184--244,
  \href{http://xxx.lanl.gov/abs/hep-th/9411210}{{\tt hep-th/9411210}}.

\bibitem{Kogut:1974ag}
J.~B. Kogut and L.~Susskind, ``{Hamiltonian Formulation of Wilson's Lattice
  Gauge Theories},'' {\em Phys. Rev.} {\bf D11} (1975) 395--408.

\bibitem{Casini:2006es}
H.~Casini and M.~Huerta, ``{A c-theorem for the entanglement entropy},'' {\em
  J. Phys.} {\bf A40} (2007) 7031--7036,
  \href{http://xxx.lanl.gov/abs/cond-mat/0610375}{{\tt cond-mat/0610375}}.

\bibitem{Kitaev:2005dm}
A.~Kitaev and J.~Preskill, ``{Topological entanglement entropy},'' {\em Phys.
  Rev. Lett.} {\bf 96} (2006) 110404,
  \href{http://xxx.lanl.gov/abs/hep-th/0510092}{{\tt hep-th/0510092}}.

\bibitem{Levin:2006zz}
M.~Levin and X.-G. Wen, ``{Detecting Topological Order in a Ground State Wave
  Function},'' {\em Phys. Rev. Lett.} {\bf 96} (2006) 110405.

\bibitem{Radicevic:2014kqa}
D.~Radicevic, ``{Notes on Entanglement in Abelian Gauge Theories},''
  \href{http://xxx.lanl.gov/abs/1404.1391}{{\tt 1404.1391}}.

\bibitem{Kabat:1995jq}
D.~N. Kabat, S.~H. Shenker, and M.~J. Strassler, ``{Black hole entropy in the
  O(N) model},'' {\em Phys. Rev.} {\bf D52} (1995) 7027--7036,
  \href{http://xxx.lanl.gov/abs/hep-th/9506182}{{\tt hep-th/9506182}}.

\bibitem{Harlow:2015lma}
D.~Harlow, ``{Wormholes, Emergent Gauge Fields, and the Weak Gravity
  Conjecture},'' {\em JHEP} {\bf 01} (2016) 122,
  \href{http://xxx.lanl.gov/abs/1510.07911}{{\tt 1510.07911}}.

\bibitem{Donnelly:2015hta}
W.~Donnelly and S.~B. Giddings, ``{Diffeomorphism-invariant observables and
  their nonlocal algebra},'' {\em Phys. Rev.} {\bf D93} (2016), no.~2 024030,
  \href{http://xxx.lanl.gov/abs/1507.07921}{{\tt 1507.07921}}. [Erratum: Phys.
  Rev.D94,no.2,029903(2016)].

\bibitem{Jafferis:2015del}
D.~L. Jafferis, A.~Lewkowycz, J.~Maldacena, and S.~J. Suh, ``{Relative entropy
  equals bulk relative entropy},'' {\em JHEP} {\bf 06} (2016) 004,
  \href{http://xxx.lanl.gov/abs/1512.06431}{{\tt 1512.06431}}.

\bibitem{Donnelly:2016qqt}
W.~Donnelly, B.~Michel, D.~Marolf, and J.~Wien, ``{Living on the Edge: A Toy
  Model for Holographic Reconstruction of Algebras with Centers},''
  \href{http://xxx.lanl.gov/abs/1611.05841}{{\tt 1611.05841}}.

\bibitem{Mcgough:2013gka}
L.~McGough and H.~Verlinde, ``{Bekenstein-Hawking Entropy as Topological
  Entanglement Entropy},'' {\em JHEP} {\bf 11} (2013) 208,
  \href{http://xxx.lanl.gov/abs/1308.2342}{{\tt 1308.2342}}.

\bibitem{Almheiri:2016blp}
A.~Almheiri, X.~Dong, and B.~Swingle, ``{Linearity of Holographic Entanglement
  Entropy},'' {\em JHEP} {\bf 02} (2017) 074,
  \href{http://xxx.lanl.gov/abs/1606.04537}{{\tt 1606.04537}}.

\bibitem{Susskind:1994sm}
L.~Susskind and J.~Uglum, ``{Black hole entropy in canonical quantum gravity
  and superstring theory},'' {\em Phys. Rev.} {\bf D50} (1994) 2700--2711,
  \href{http://xxx.lanl.gov/abs/hep-th/9401070}{{\tt hep-th/9401070}}.

\bibitem{Jensen:2013ora}
K.~Jensen and A.~Karch, ``{Holographic Dual of an Einstein-Podolsky-Rosen Pair
  has a Wormhole},'' {\em Phys. Rev. Lett.} {\bf 111} (2013), no.~21 211602,
  \href{http://xxx.lanl.gov/abs/1307.1132}{{\tt 1307.1132}}.

\bibitem{Lewkowycz:2013laa}
A.~Lewkowycz and J.~Maldacena, ``{Exact results for the entanglement entropy
  and the energy radiated by a quark},'' {\em JHEP} {\bf 05} (2014) 025,
  \href{http://xxx.lanl.gov/abs/1312.5682}{{\tt 1312.5682}}.

\bibitem{Dong:2016fnf}
X.~Dong, ``{The Gravity Dual of Renyi Entropy},'' {\em Nature Commun.} {\bf 7}
  (2016) 12472, \href{http://xxx.lanl.gov/abs/1601.06788}{{\tt 1601.06788}}.

\bibitem{Czech:2012be}
B.~Czech, J.~L. Karczmarek, F.~Nogueira, and M.~Van~Raamsdonk, ``{Rindler
  Quantum Gravity},'' {\em Class. Quant. Grav.} {\bf 29} (2012) 235025,
  \href{http://xxx.lanl.gov/abs/1206.1323}{{\tt 1206.1323}}.

\bibitem{Donnelly:2016jet}
W.~Donnelly and G.~Wong, ``{Entanglement branes in a two-dimensional string
  theory},'' \href{http://xxx.lanl.gov/abs/1610.01719}{{\tt 1610.01719}}.

\bibitem{Baez:1994gk}
J.~Baez and W.~Taylor, ``{Strings and two-dimensional QCD for finite N},'' {\em
  Nucl. Phys.} {\bf B426} (1994) 53--70,
  \href{http://xxx.lanl.gov/abs/hep-th/9401041}{{\tt hep-th/9401041}}.

\bibitem{Gross:1993yt}
D.~J. Gross and W.~Taylor, ``{Twists and Wilson loops in the string theory of
  two-dimensional QCD},'' {\em Nucl. Phys.} {\bf B403} (1993) 395--452,
  \href{http://xxx.lanl.gov/abs/hep-th/9303046}{{\tt hep-th/9303046}}.

\bibitem{Witten:1991yr}
E.~Witten, ``{On string theory and black holes},'' {\em Phys. Rev.} {\bf D44}
  (1991) 314--324.

\bibitem{Giveon:2013ica}
A.~Giveon and N.~Itzhaki, ``{String theory at the tip of the cigar},'' {\em
  JHEP} {\bf 09} (2013) 079, \href{http://xxx.lanl.gov/abs/1305.4799}{{\tt
  1305.4799}}.

\bibitem{Giveon:2014hfa}
A.~Giveon, N.~Itzhaki, and J.~Troost, ``{Lessons on Black Holes from the
  Elliptic Genus},'' {\em JHEP} {\bf 04} (2014) 160,
  \href{http://xxx.lanl.gov/abs/1401.3104}{{\tt 1401.3104}}.

\bibitem{Giveon:2015raa}
A.~Giveon, J.~Harvey, D.~Kutasov, and S.~Lee, ``{Three-Charge Black Holes and
  Quarter BPS States in Little String Theory},'' {\em JHEP} {\bf 12} (2015)
  145, \href{http://xxx.lanl.gov/abs/1508.04437}{{\tt 1508.04437}}.

\bibitem{Troost:2017fpk}
J.~Troost, ``{An Elliptic Triptych},''
  \href{http://xxx.lanl.gov/abs/1706.02576}{{\tt 1706.02576}}.

\bibitem{Hartnoll:2015fca}
S.~A. Hartnoll and E.~Mazenc, ``{Entanglement entropy in two dimensional string
  theory},'' {\em Phys. Rev. Lett.} {\bf 115} (2015), no.~12 121602,
  \href{http://xxx.lanl.gov/abs/1504.07985}{{\tt 1504.07985}}.

\bibitem{Karczmarek:2004bw}
J.~L. Karczmarek, J.~M. Maldacena, and A.~Strominger, ``{Black hole
  non-formation in the matrix model},'' {\em JHEP} {\bf 01} (2006) 039,
  \href{http://xxx.lanl.gov/abs/hep-th/0411174}{{\tt hep-th/0411174}}.

\bibitem{Kazakov:2000pm}
V.~Kazakov, I.~K. Kostov, and D.~Kutasov, ``{A Matrix model for the
  two-dimensional black hole},'' {\em Nucl. Phys.} {\bf B622} (2002) 141--188,
  \href{http://xxx.lanl.gov/abs/hep-th/0101011}{{\tt hep-th/0101011}}.

\bibitem{Ohya}
M.~Ohya and D.~Petz, {\em Quantum Entropy and its Use}.
\newblock Springer, 2004.

\bibitem{Radicevic:2016tlt}
D.~Radi{\v c}evi{\'c}, ``{Entanglement Entropy and Duality},'' {\em JHEP} {\bf
  11} (2016) 130, \href{http://xxx.lanl.gov/abs/1605.09396}{{\tt 1605.09396}}.

\bibitem{Donnelly:2016mlc}
W.~Donnelly, B.~Michel, and A.~Wall, ``{Electromagnetic Duality and
  Entanglement Anomalies},'' \href{http://xxx.lanl.gov/abs/1611.05920}{{\tt
  1611.05920}}.

\bibitem{Kitaev:1997wr}
A.~{\relax Yu}. Kitaev, ``{Fault tolerant quantum computation by anyons},''
  {\em Annals Phys.} {\bf 303} (2003) 2--30,
  \href{http://xxx.lanl.gov/abs/quant-ph/9707021}{{\tt quant-ph/9707021}}.

\bibitem{Wen:2004ym}
X.~G. Wen, {\em QFT of many-body systems: From the origin of sound to an origin
  of light and electrons}.
\newblock 2004.

\bibitem{McGreevy:2016myw}
J.~McGreevy, ``{TASI lectures on quantum matter (with a view toward holographic
  duality)},'' in {\em {Proceedings, TASI 2015}}, pp.~215--296, 2017.
\newblock \href{http://xxx.lanl.gov/abs/1606.08953}{{\tt 1606.08953}}.

\bibitem{Lee:2010fy}
S.-S. Lee, ``{TASI Lectures on Emergence of Supersymmetry, Gauge Theory and
  String in Condensed Matter Systems},'' in {\em {Proceedings, TASI 2010}},
  pp.~667--706, 2010.
\newblock \href{http://xxx.lanl.gov/abs/1009.5127}{{\tt 1009.5127}}.

\end{thebibliography}\endgroup


\begingroup\raggedright\endgroup

\end{document}